\documentclass[superscriptaddress,aps,preprintnumbers,amsmath,showpacs,amssymb,prl,nofootinbib,reprint]{revtex4-1}
\usepackage{bm, color} 
\usepackage{amssymb,amsfonts,slashed,amsthm,amsmath,graphicx, soul}
\usepackage[caption=false]{subfig}
\usepackage{hyperref}
\usepackage{tikz,xcolor}

\begin{document}


\newcommand{\vev}[1]{ \left\langle {#1} \right\rangle }
\newcommand{\bra}[1]{ \langle {#1} | }
\newcommand{\ket}[1]{ | {#1} \rangle }
\newcommand{\eV}{ \ {\rm eV} }
\newcommand{\KeV}{ \ {\rm keV} }
\newcommand{\MeV}{\  {\rm MeV} }
\newcommand{\GeV}{\  {\rm GeV} }
\newcommand{\TeV}{\  {\rm TeV} }
\newcommand{\1}{\mbox{1}\hspace{-0.25em}\mbox{l}}
\newcommand{\Red}[1]{{\color{red} {#1}}}

\newcommand{\lmk}{\left(}  
\newcommand{\rmk}{\right)}
\newcommand{\lkk}{\left[}  
\newcommand{\rkk}{\right]}
\newcommand{\lhk}{\left \{ }  
\newcommand{\rhk}{\right \} }
\newcommand{\del}{\partial}  
\newcommand{\la}{\left\langle} 
\newcommand{\ra}{\right\rangle}
\newcommand{\half}{\frac{1}{2}}

\newcommand{\bea}{\begin{array}}
\newcommand{\eea}{\end{array}}
\newcommand{\beq}{\begin{eqnarray}}
\newcommand{\eeq}{\end{eqnarray}}
\newcommand{\eq}[1]{Eq.~(\ref{#1})}

\newcommand{\dd}{\mathrm{d}}
\newcommand{\Mpl}{M_{\rm Pl}}
\newcommand{\mg}{m_{3/2}}
\newcommand{\abs}[1]{\left\vert {#1} \right\vert}
\newcommand{\mphi}{m_{\phi}}
\newcommand{\Hz}{\ {\rm Hz}}
\newcommand{\for}{\quad \text{for }}
\newcommand{\Min}{\text{Min}}
\newcommand{\Max}{\text{Max}}
\newcommand{\Kahler}{K\"{a}hler }
\newcommand{\cphi}{\varphi}
\newcommand{\Tr}{\text{Tr}}
\newcommand{\diag}{{\rm diag}}

\newcommand{\SUf}{SU(3)_{\rm f}}
\newcommand{\Upq}{U(1)_{\rm PQ}}
\newcommand{\Zpq}{Z^{\rm PQ}_3}
\newcommand{\Cpq}{C_{\rm PQ}}
\newcommand{\ubar}{u^c}
\newcommand{\dbar}{d^c}
\newcommand{\ebar}{e^c}
\newcommand{\nubar}{\nu^c}
\newcommand{\Ndw}{N_{\rm DW}}
\newcommand{\Fpq}{F_{\rm PQ}}
\newcommand{\fpq}{v_{\rm PQ}}
\newcommand{\Br}{{\rm Br}}
\newcommand{\Lag}{\mathcal{L}}
\newcommand{\Lqcd}{\Lambda_{\rm QCD}}

\newcommand{\ji}{j_{\rm inf}} 
\newcommand{\jb}{j_{B-L}} 
\newcommand{\M}{M} 
\newcommand{\im}{{\rm Im} }
\newcommand{\re}{{\rm Re} }

\def\lrf#1#2{ \left(\frac{#1}{#2}\right)}
\def\lrfp#1#2#3{ \left(\frac{#1}{#2} \right)^{#3}}
\def\lrp#1#2{\left( #1 \right)^{#2}}
\def\REF#1{Ref.~\cite{#1}}
\def\SEC#1{Sec.~\ref{#1}}
\def\FIG#1{Fig.~\ref{#1}}
\def\EQ#1{Eq.~(\ref{#1})}
\def\EQS#1{Eqs.~(\ref{#1})}
\def\TEV#1{10^{#1}{\rm\,TeV}}
\def\GEV#1{10^{#1}{\rm\,GeV}}
\def\MEV#1{10^{#1}{\rm\,MeV}}
\def\KEV#1{10^{#1}{\rm\,keV}}
\def\blue#1{\textcolor{blue}{#1}}
\def\red#1{\textcolor{red}{#1}}

\newcommand{\eff}{\Delta N_{\rm eff}}
\newcommand{\neff}{\Delta N_{\rm eff}}
\newcommand{\cc}{\Lambda}
\newcommand{\Mpc}{\ {\rm Mpc}}
\newcommand{\Msolar}{M_\cdot}


\preprint{TU-1118,~IPMU21-0018}

\title{
Cosmic Birefringence Triggered by Dark Matter Domination
}

\author{
Shota Nakagawa
}
\affiliation{Department of Physics, Tohoku University, 
Sendai, Miyagi 980-8578, Japan} 

\author{
Fuminobu Takahashi
}
\affiliation{Department of Physics, Tohoku University, 
Sendai, Miyagi 980-8578, Japan} 
\affiliation{Kavli IPMU (WPI), UTIAS, The University of Tokyo, 
Kashiwa, Chiba 277-8583, Japan} 

\author{
Masaki Yamada
}
\affiliation{Department of Physics, Tohoku University, 
Sendai, Miyagi 980-8578, Japan} 
\affiliation{Frontier Research Institute for Interdisciplinary Sciences, Tohoku University, 
Sendai, Miyagi 980-8578, Japan}

\begin{abstract}
Cosmic birefringence is predicted if an axion-like particle (ALP) moves after the recombination. We show that this naturally happens if the ALP is coupled to the dark matter density because it then acquires a large effective mass after the matter-radiation equality. Our scenario applies to a broad range of the ALP mass $m_\phi \lesssim 10^{-28}$\,eV, even smaller than the present Hubble constant. We give a simple model to realize this scenario, where dark matter is made of hidden monopoles, which give the ALP such a large effective mass through the Witten effect. The mechanism works if the ALP decay constant is of order the GUT scale without a fine-tuning of the initial misalignment angle. For smaller decay constant, the hidden monopole can be a fraction of dark matter. We also study the implications for the QCD axion, and show that the domain wall problem can be solved by the effective mass.
\end{abstract}

\maketitle
\flushbottom

{\bf Introduction.--}
Axions are ubiquitous in string theory and are known to have interesting effects on various observables, despite the fact that their interactions are suppressed by large mass scales such as the string scale.
For instance, the collective excitations, like coherent oscillations, can be induced during the evolution of the universe~\cite{Preskill:1982cy,Abbott:1982af,Dine:1982ah}, which may explain all or part of dark matter (DM). A large number of axions are also produced by the decay of moduli fields, contributing to the effective number of neutrino species~\cite{Ichikawa:2007jv,Cicoli:2012aq,Higaki:2012ar,Higaki:2013lra,Cicoli:2018cgu,Takahashi:2019ypv}. 
In addition to gravitational interactions,
axions can have interactions with the Standard Model particles.  Among them, an anomalous coupling with photons is  known to cause very diverse phenomena in cosmology and astrophysics, such as X- or $\gamma$-ray emission from the decay of axions~\cite{Cadamuro:2011fd,Arias:2012az,Higaki:2014zua,Jaeckel:2014qea,Higaki:2014qua} (\cite{Kawasaki:1997ah,Asaka:1997rv,Asaka:1999xd} for early works on a similar subject), 
superradiance around  rotating black holes~\cite{Arvanitaki:2014wva,Cardoso:2018tly,Stott:2018opm,Davoudiasl:2019nlo,Sun:2019mqb,Palomba:2019vxe}, and cosmic birefringence (CB), i.e. the rotation of photon polarization plane~\cite{Carroll:1989vb,Carroll:1991zs, Harari:1992ea,Carroll:1998zi,Lue:1998mq,Pospelov:2008gg,Fedderke:2019ajk,Agrawal:2019lkr}.
Axions coupled to photons are often referred to as axion-like particles (ALPs).

Recently, it was reported in Ref.~\cite{Minami:2020odp} that 
the Planck 2018 polarization data of cosmic microwave background (CMB) favors a nonzero value of isotropic CB,
with statistical significance of 2.4 $\sigma$, based on the novel method~\cite{Minami:2019ruj,Minami:2020xfg,Minami:2020fin}. The suggested rotation angle of the CMB polarization is\footnote{
A positive $\beta$ corresponds to clockwise rotation for an observer.
}
\begin{align}
\label{obsCB}
    \beta= 0.35 \pm 0.14 {\rm~ deg },
\end{align}
and future observations of CMB will reduce statistical uncertainties by more than one order of magnitude~\cite{Pogosian:2019jbt}. One plausible explanation of the isotropic CB is an ALP which starts to move during or after the recombination epoch.\footnote{
Another possibility is the axion domain wall, which works for heavier ALP masses, and predicts peculiar anisotropic CB~\cite{Takahashi:2020tqv}.
} The ALP mass relevant for this scenario is
 in the range of $(10^{-33}\eV,10^{-28} \eV)$, because, for heavier or lighter ALP masses,
 a large enhancement of the ALP-photon coupling is required~\cite{Fujita:2020aqt}.

Interestingly,  the isotropic CB implied by the current observation (\ref{obsCB})
is very natural from a theoretical point of view. It is determined by the 
ratio of the change in the ALP 
to its decay constant, and is not sensitive to the decay constant itself. 
The observed rotation angle suggests that
the ALP has changed by about the decay constant from the time of recombination to the present.
In other words, no fine-tuning of the initial value is necessary. 
Even so, one might wonder why the ALP has started moving at such a
special timing in the history of the universe, namely, between the recombination and the present. The timing of when the ALP begins to oscillate is determined by the balance between the Hubble parameter and the ALP mass, which requires the ALP mass to be in the mass range described above. This may seem ad hoc, if there were not for any particular theoretical reason. One might think that, if the ALP masses are logarithmically distributed on each scale as suggested in the string axiverse~\cite{Arvanitaki:2014wva}, this is naturally explained (see \cite{Mehta:2021pwf} for a recent analysis).
However, it is by no means obvious that this is indeed the case, since it is not known how many axions actually exist in the low energy and whether they are coupled to photons in the Standard Model. In fact, if the string axion masses are logarithmically distributed, the smallness of the mass itself may not be a problem, but it actually implies that there is no natural lower bound on the axion mass. Therefore, it can still be regarded as a  ``coincidence", i.e. some kind of fine-tuning problem regarding the timing of the ALP oscillations.

In this letter, we present a scenario that explains the reason for the ``coincidence" of 
the recent ALP oscillations. Our scenario is based on the observational fact that  the recombination and matter-radiation equality occur in close proximity. We point out that if the ALP starts to move via coupling with the DM density after the matter-radiation equality, it naturally induces the isotropic CB (\ref{obsCB}) for a broader range of the mass. From the point of view of low-energy effective field theory, such a phenomenon occurs if the ALP acquires an effective mass of the order of the Hubble parameter during the matter-dominated era. Such an effective mass may come from the  interaction between the ALP and (dark) matter with gravitational strength.  We will present a concrete model in which the ALP acquires an effective mass via the Witten effect of hidden monopole DM. We discuss that,
if the QCD axion acquires a similar effective mass,
the isocurvature and domain wall problems can also be solved.

\vspace{2mm}
{\bf Cosmic birefringence from ALP dynamics.--}
We consider an (almost) massless ALP that couples to photons via anomaly: 
\beq
 {\cal L} 
 \supset -c_\gamma \frac{\alpha}{4 \pi} \frac{\phi}{f_\phi} F_{\mu \nu} \tilde{F}^{\mu \nu} 
 \equiv -\frac{1}{4} g_{\phi \gamma \gamma} \phi F_{\mu \nu} \tilde{F}^{\mu \nu}, 
\eeq
where $\alpha$ is the fine-structure constant, 
$f_\phi$  the ALP decay constant, $c_\gamma$  the U(1)$_{\rm EM}$ anomaly coefficient, 
and $F_{\mu \nu}$ and $\tilde{F}^{\mu \nu} \equiv \epsilon^{\mu\nu\rho\sigma}F_{\rho\sigma}/2\sqrt{-g}$ 
the field strength and its dual. 
In terms of the electric and magnetic fields we have  $F_{\mu \nu} \tilde{F}^{\mu \nu} = -4 {\bm E} \cdot {\bm B}$. 
The natural values of $c_\gamma$ is of ${\cal O}(1)$, but it can be much larger than unity
in a contrived set-up~\cite{Higaki:2016yqk,Farina:2016tgd}.
For simplicity, we assume that 
the ALP mass is smaller than the current Hubble constant $H_0\simeq 10^{-33}\eV$,
and neglect the mass in the following, although it can be straightforwardly extended to $m_\phi \lesssim 10^{-28}\eV$.
As we will see, our scenario not only extends the viable mass region to smaller masses, but also easily satisfies the upper limit on the  ALP abundance~\cite{Hlozek:2014lca}.
The rotation angle of the CMB polarization is related 
to the change of the ALP field value
from the LSS to the present, $\Delta \phi \equiv \phi_{\rm p}-\phi_{\rm LSS}$, as~\cite{Harari:1992ea}\footnote{
Note that there is a sign error after their Eq. (4) of \cite{Harari:1992ea}.
}
\beq
 \beta \simeq 0.42 {\rm~ deg}\, \times 
 \lmk
 c_\gamma \frac{\Delta \phi}{2 \pi f_\phi} \rmk.
 \label{obs}
\eeq

Now we introduce an effective Hubble-induced ALP mass from 
the DM density: 
\beq
 V(\phi) = \frac12 c_H H_{\rm DM}^2 \phi^2, 
 \label{V}
\eeq
where we assume $c_H$ is a positive constant and 
\beq
 H_{\rm DM}^2 \equiv \frac{\rho_{\rm DM}}{3 \Mpl^2},
\eeq
with $\rho_{\rm DM}$ being the energy density of DM, and $\Mpl$ the reduced Planck mass. Here, without loss of generality, the potential minimum
 $\phi_{\rm min}$  of the effective potential is set to be at the origin.
The Witten effect on the ALP potential in a monopole DM model is actually written in this form as we will see shortly. 
Alternatively, one may introduce a coupling to the Ricci scalar $R$ ($=6[(\dot{a}/a)^2 + \ddot{a}/a]$): 
\beq
 {\cal L} \supset - \xi 
 R \phi^2, 
\eeq
where $\xi = \mathcal{O}(1)$ is a positive constant. 
While the effective mass from this operator is negligibly small 
during the radiation-dominated era 
because of the conformal symmetry, it is about $\sqrt{6 \xi} H$ during the matter-dominated era. 
This term therefore has a similar effect as \eq{V}~\cite{Takahashi:2015waa}. 
In the following we use \eq{V} as an effective potential for the ALP.

We consider a homogeneous ALP in the following, and we will later comment on its quantum fluctuations generated during inflation.
The equation of motion for the homogeneous mode is given by 
\beq
 \ddot{\phi} + 3 H \dot{\phi} + c_H H_{\rm DM}^2 \phi = 0, 
 \label{eom}
\eeq
where the dot represents a derivative with respect to time 
and $H$ is the Hubble parameter obeying 
\beq
H^2=H^2_0\left(\Omega_{\rm{rad}}a^{-4}+\Omega_{\rm{mat}}a^{-3}+\Omega_\Lambda\right),
\eeq
with the present value of the scale factor 
 $a_0$ set to be unity.
The effective mass is negligible until the matter-radiation equality because $H_{\rm DM} \ll H$ during the radiation-dominated era. Then, the ALP starts to move toward the potential minimum after the equality, and it experiences damped oscillations
since the effective mass is comparable to the Hubble parameter. 
As a result, the ALP abundance can be negligibly small at present. 
Since the recombination occurs soon after the matter-radiation equality, 
a nonzero amount of isotropic CB is induced.

\vspace{2mm}
{\bf Numerical results.--}
We numerically solve the equation of motion (\ref{eom}) with various values of $c_H$ and initial conditions. 
We denote the initial ALP field value as $\phi_i$ and the potential minimum as $\phi_{\rm min}(=0)$, its present field value as $\phi_{\rm p}$, and the corresponding angle, $\theta = \phi/f_\phi$, with the same indices. Note that $\phi_i$ is not necessarily equal to $\phi_{\rm LSS}$ especially for $c_H \gg 1$, since the ALP may start oscillating before recombination. 
Similarly, $\phi_{\rm p}$ is not necessarily equal to $\phi_{\rm min}$ especially for $c_H \ll 1$, since the mass may be too small for the ALP  to settle down at the potential minimum. 
Since the ALP moves toward the origin until present, we have ${\rm sign}[\Delta \phi] = {\rm sign}[\phi_{\rm min} - \phi_i]$ at least for $c_H \lesssim 1$.
Thus, for a positive $c_\gamma$, we should take $\phi_i < 0$ so as to be consistent with the observational result (\ref{obs}).

In our scenario the ALP tends to start moving slightly before the recombination because the effective mass becomes relevant after the matter-radiation equality. 
We thus need to take account of the thickness of LSS.
We estimate the ALP field value at the LSS by weighing it with the visibility function $g(T)$\cite{Weinberg:2008zzc},
\beq
\la \phi_{\rm{LSS}} \ra =\int dT g(T)\phi(T).
\eeq
Then, $\beta$ is given by \eq{obs} with $\Delta \phi = \phi_{\rm p} - \la \phi_{\rm LSS} \ra$.

\begin{figure}[t!]
\includegraphics[width=8cm]{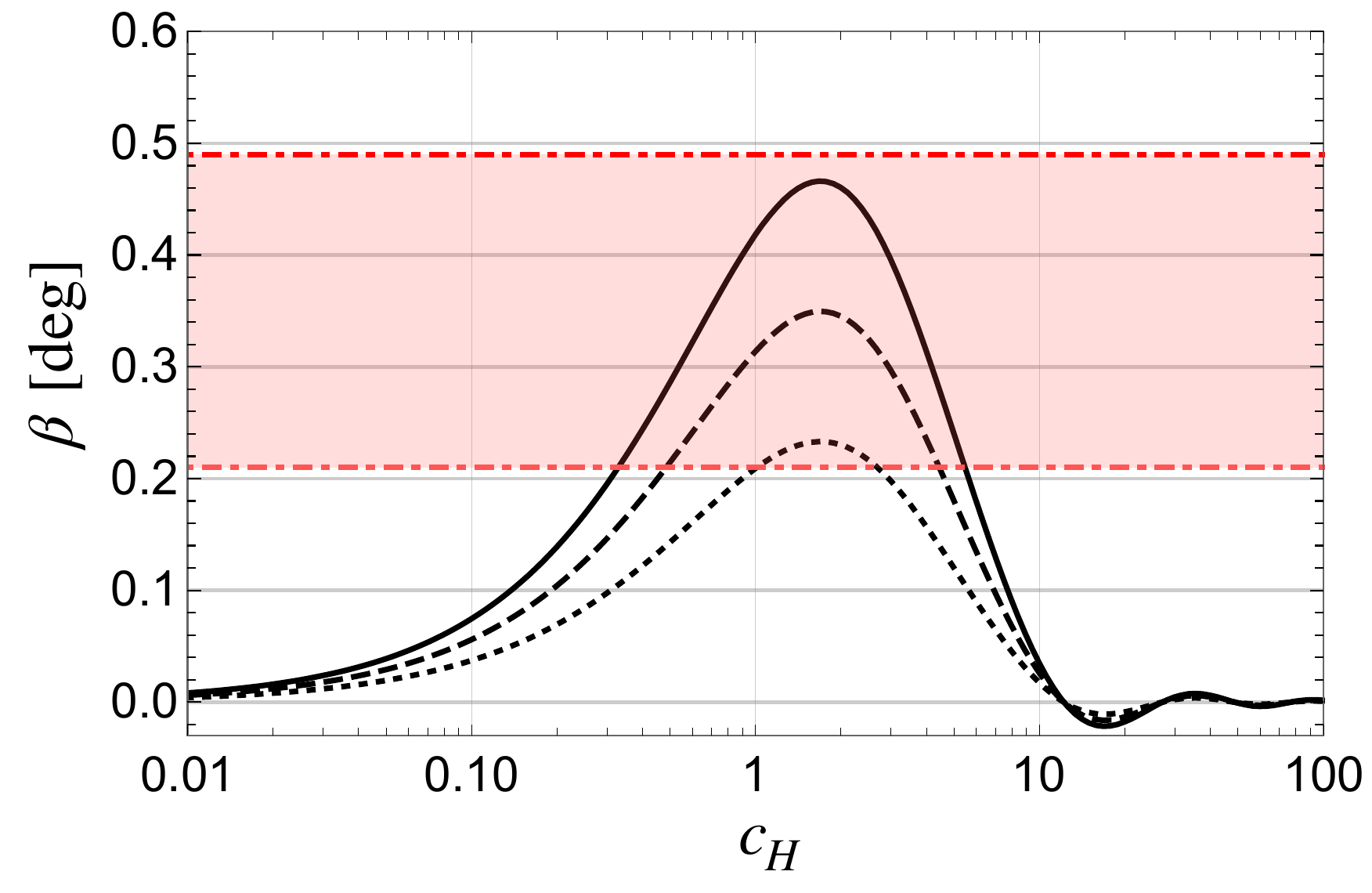}
\centering
\caption{
The predicted rotation angle $\beta$ as a function of $c_H$.
The solid, dashed, and dotted black lines correspond to $c_\gamma=12, 9, 6$, from top to bottom, respectively.
We take the initial condition, $(\theta_i, \dot{\theta}_i)=(-1,0)$.
The shaded region shows the region favored by the Planck polarization data~(\ref{obsCB}).
}
\label{fig:result}
\end{figure}

The resulting rotation angle $\beta$ is shown 
as a function of $c_H$ in Fig.~\ref{fig:result}.
The solid, dashed, and dotted black lines represent the cases with $c_\gamma=12, 9, 6$ from top to bottom, and the red shaded region shows the allowed region by the Planck polarization data (\ref{obsCB}).
We take the initial condition $(\theta_i, \dot{\theta}_i)=(-1,0)$ as an example. 
One can rescale our result for a different $\theta_i$ by using $\beta \propto (\theta_{\rm min}- \theta_i)$, since the equation of motion  \eq{eom} is linear in $\phi$.\footnote{
This is a good approximation even for a cosine function instead of the quadratic term, as long as the anharmonic effect is small.
}
As a result, 
the observational hint for the isotropic CB can be explained 
if $c_\gamma (\theta_{\rm min}-\theta_i) \gtrsim 6$ for $c_H = \mathcal{O}(1)$.

The qualitative behavior can be understood as follows. 
For $c_H \ll 1$, the effective mass is too small and the ALP does not move much even during the matter-dominated era. On the contrary, for $c_H \gg 1$, the ALP starts to oscillate well before the matter-radiation equality and it already settles down at the potential minimum by the recombination. Because of the balance between these effects, the rotation angle is maximum at $c_H \sim 2$. 
For a moderately large $c_H=\mathcal{O}(10)$, the ALP oscillates a few times around the potential minimum by the recombination and can have an opposite sign at the LSS. This oscillatory behavior can be  seen in Fig.~\ref{fig:result}.

\begin{figure}[t!]
\includegraphics[width=8cm]{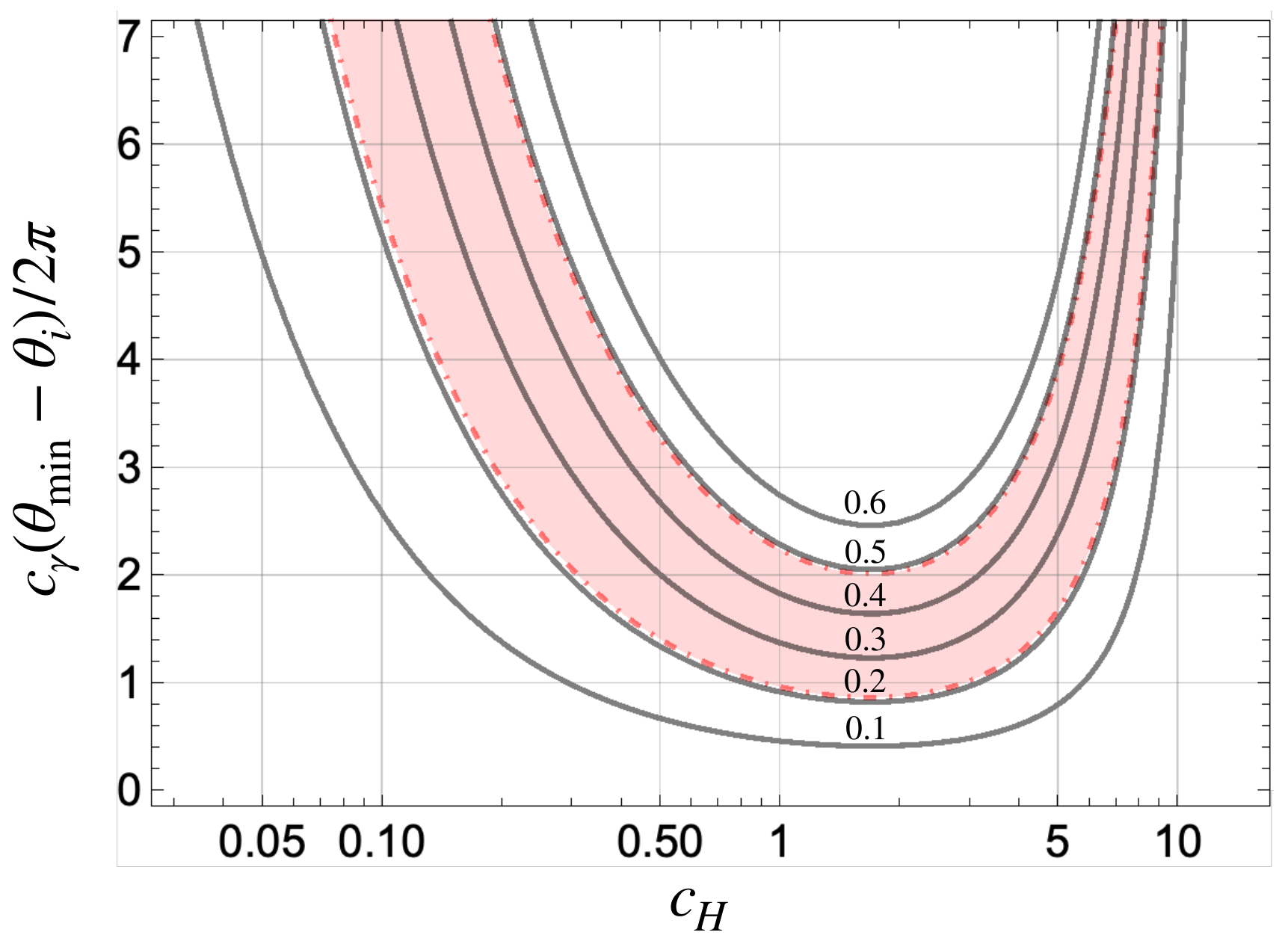}
\centering
\caption{
The contour plot of the rotation angle $\beta$ as a function of $c_H$ and $c_\gamma(\theta_{\rm{min}}-\theta_{i})/2\pi$.
The shaded region shows the observational hint
for the isotropic CB~(\ref{obsCB}).
}
\label{fig:contour}
\end{figure}

In Fig.~\ref{fig:contour} we show a contour of the rotation angle $\beta$ as a function of  $c_H$ and $c_\gamma (\theta_{\rm min} - \theta_i) / (2 \pi)$. 
For $c_H = \mathcal{O}(1)$, we need $c_\gamma (\theta_{\rm min} - \theta_i) / (2 \pi) = \mathcal{O}(1)$ to obtain the observed value (\ref{obsCB}). In other words, no fine-tuning of the initial condition is required in this case, for $c_\gamma = {\cal O}(1)$.
We can see that larger values of $\theta_{\rm min}-\theta_i$ and/or $c_\gamma$ are required for larger or smaller values of $c_H$.

\vspace{2mm}
{\bf Effective mass from the Witten effect.--}
Now we provide an explicit model in which the ALP acquires the effective Hubble-induced mass via the Witten effect from hidden monopole DM. 
A monopole associated with a hidden U(1)$_H$ is 
stable due to its magnetic charge, and is therefore
a good candidate for DM.
If the ALP has an anomalous coupling to the U(1)$_H$ photons, we
obtain the aforementioned coupling between the ALP and DM density as in \eq{V}.

First let us see the effect of the $\Theta$-term
on monopoles.
The Lagrangian of the hidden U(1)$_H$ gauge field is given by 
\beq
\mathcal{L}\supset-\frac{1}{4}X_{\mu\nu}X^{\mu\nu}-\frac{\alpha_H\Theta}{8\pi}X_{\mu\nu}\tilde{X}^{\mu\nu}
\eeq
where $\alpha_H$ is a fine-structure constant for U(1)$_H$, $\Theta$ is a CP violating angle, 
and $X_{\mu\nu}\equiv\del_\mu X_\nu-\del_\nu X_\mu$ denotes the field strength of the U(1)$_H$ gauge field $X_\mu$.
Although the second term is a total derivative, it modifies the Maxwell equation as 
\beq
\nabla\cdot\bm{E}_{H}=-\frac{\alpha_H}{2\pi}\nabla\cdot(\Theta\bm{B}_H),
\label{Maxwell}
\eeq
where $(\bm{E}_H)_i\equiv X_{0i}$, $(\bm{B}_H)_i\equiv-\epsilon_{ijk}X^{jk}/2$.
A magnetic monopole with a magnetic charge $g_H$ sitting at the origin generates a magnetic field, satisfying the Gauss's law, $\nabla\cdot\bm{B}_H=g_H \delta^{(3)}(0)$.
The modified Maxwell equation (\ref{Maxwell})  implies that the magnetic monopole acquires an electric charge proportional to the $\Theta$ parameter, and becomes a dyon. This is known as the Witten effect~\cite{Witten:1979ey}.

Now we introduce the ALP with an anomalous coupling to the U(1)$_H$ gauge field. To this end we promote $\Theta$ to the ALP field  by the replacement of $\Theta \to \phi / f_\phi$. 
According to \eq{Maxwell}, a nonzero field value of $\phi$ induces an electric field around the monopole. This means that the total electric energy around a monopole $V_M$ depends on $\phi$
as
\beq
V_M\simeq
\frac{\alpha_H}{32\pi^2 r_c}
\frac{\phi^2}{f_\phi^2},
\eeq
where $r_c$ is the radius of the monopole core.
Considering the 't Hooft-Polyakov monopole \cite{tHooft:1974kcl,Polyakov:1974ek}, we have $r_c\sim m_{W_H}^{-1}$. Here $m_{W_H}$ is the mass of a heavy gauge field, which is about 
the monopole mass $m_M$ multiplied by $\alpha_H$.
See Refs.~\cite{Murayama:2009nj,Baek:2013dwa, Khoze:2014woa} for the production and abundances
of the hidden monopole and heavy gauge bosons.
Taking the spatial average, we obtain the energy density of the ALP ground state as $V(\phi)=n_M V_M(\phi)$ where $n_M\equiv n_{M+}+ n_{M-}$ denotes the number density of monopoles and anti-monopoles.
Thus, the ALP potential is given in the form of (\ref{V}) with
\beq
 c_H = 
 3 \lmk \frac{\rho_M}{\rho_{\rm DM}} \rmk
 \lmk \frac{\alpha_H}{4 \pi} \frac{\Mpl}{f_\phi}
 \rmk^2,
 \label{Witten}
\eeq
where $\rho_M$ denotes the energy density of the monopole~\cite{Fischler:1983sc}.

Before proceeding, let us comment on an upper bound on $\alpha_H$ in our analysis. 
We expect that monopoles appear from spontaneous symmetry breaking of some non-Abelian gauge theory, such as SU(2)$_H$. 
Then the instanton effect of SU$(2)_H$ gives rise to an additional mass term about $\phi=0$, though it is suppressed by the instanton exponent, $\sim e^{-\pi/\alpha_H}$ \cite{Fuentes-Martin:2019bue,Csaki:2019vte,Buen-Abad:2019uoc}.
This additional mass is negligible and the ALP is approximately massless (except the effective mass \eq{V} via the Witten effect) when $\alpha_H\lesssim 0.02$. Although our analysis can be straightforwardly extended to a larger gauge coupling
if $m_\phi \lesssim 10^{-28}\eV$, we assume this inequality for simplicity.

As we have seen above, $c_H = \mathcal{O}(1)$ is needed to explain the isotropic CB \eq{obsCB}. This is realized for the ALP decay constant of the order of the GUT scale ($\approx 10^{15\,\text{-}\,16}\GeV$) and $\alpha_H \sim \alpha$. To be explicit,
for $\alpha_H=0.02$, $\rho_M=\rho_{\rm{DM}}$, and $f_\phi=5 \times 10^{15}\GeV$, we obtain $c_H\simeq 2$ from \eq{Witten}.
This shows that the ALP coupled to hidden monopole DM can generate the isotropic CB without a fine-tuning of the initial misalignment, and the GUT-scale decay constant suggests that such an ALP may be one of the string axions.

\vspace{2mm}
{\bf Connection to the QCD axion.--}
If the QCD axion is also coupled to the hidden photons, 
it acquires the effective mass given by Eq.~(\ref{V}) 
with the replacement of $c_H$ by $c_{H,a}$,\footnote{To be precise, it is a combination of the QCD axion and ALP that acquires mass from the Witten effect, but if $f_a \ll f_\phi$, it is mostly the QCD axion before the QCD phase transition.} which is given by 
\beq
 c_{H,a} = 
  3 \lmk \frac{\rho_M}{\rho_{\rm DM}} \rmk 
  \lmk \frac{N_{\rm H}}{N_{\rm DW}} \rmk^2
 \lmk \frac{\alpha_H}{4 \pi} \frac{\Mpl}{f_a}
 \rmk^2, 
\eeq
where $f_a$ is the decay constant for the QCD axion and $N_{\rm DW}$ and $N_{\rm H}$ are domain wall numbers of the QCD axion associated with SU(3)$_c$ and U(1)$_H$, respectively. 
The potential minimum of this term is generically different from the one for the QCD vacuum. 
The effective mass \eq{V} is negligible at present and does not spoil the success of the Peccei-Quinn (PQ) mechanism~\cite{Peccei:1977hh, Peccei:1977ur}, while it is stronger in the early universe and can affect the evolution of the QCD axion.

In the pre-inflationary PQ symmetry breaking scenario, 
the effective mass can be used to avoid the isocurvature problem or suppress the QCD axion abundance~\cite{Kawasaki:2015lpf,Nomura:2015xil,Kawasaki:2017xwt,Sato:2018nqy,Nakagawa:2020zjr}. However, since the hidden gauge coupling considered here is smaller than in the previous studies, the Witten effect might become relevant only for small $f_a \lesssim 10^{11}\GeV$.

In the post-inflationary PQ symmetry breaking scenario, 
the effective mass can be used to solve the domain wall problem~\cite{Kawasaki:2015lpf,Sato:2018nqy}. 
First, we note that cosmic strings form at the time of the PQ symmetry breaking when the temperature is expected to be of order $f_a$. 
Then if the effective mass due to the Witten effect
becomes larger than the Hubble parameter before the QCD phase transition, 
each cosmic string will be attached by a single domain wall if $N_H=1$. The cosmic strings therefore soon disappear due to the tension of the domain wall. 
This sets an approximately homogeneous initial condition on the QCD axion at the QCD phase transition, in which case domain walls do not form when the QCD non-perturbative effect turns on even if $N_{\rm DW} > 1$.
Thus we require $c_{H,a} H_{\rm DM}^2 \gtrsim H^2$ at the QCD phase transition, namely, 
\beq
 c_{H,a} \gtrsim 10^{8}. 
\eeq
This can be satisfied if $f_a \ll f_\phi$ since 
$c_{H,a} \sim (f_\phi/f_a)^2 c_H$; for $c_H = \mathcal{O}(1)$, $f_\phi = 10^{15\,\text{-}\,16} \GeV$, and $f_a = 10^{11} \GeV$, we have $c_{H,a} =\mathcal{O}(10^{8\,\text{-}\,10})$.
Therefore, the domain wall problem for the QCD axion can be simultaneously solved by the Witten effect. 
Note that the abundance of QCD axion in this case is given by the sum of the contributions from the decay of domain wall due to the effective mass 
and the coherent oscillation at the QCD phase transition.

Finally we comment on the mixing of the ALP and the QCD axion. 
Since the ALP $\phi$ is almost massless and the QCD axion is much heavier, the latter is identified with
the combination coupled to SU(3)$_c$ and the former  is the one orthogonal to it in the mass-eigenstate basis~\cite{Pospelov:2008gg,Takahashi:2020tqv}. In the above argument we have not considered the mixing, which can be justified for $f_a \ll f_\phi$. After the QCD phase transition, 
we have the ultralight ALP with the Witten effect by integrating out the QCD axion as long as the combination coupled to (hidden) photons is different from that to gluons.

\vspace{2mm}
{\bf Discussion and Conclusions.--}
So far we have focused on the isotropic CB, but anisotropic CB is also generated if the ALP acquires quantum fluctuations $\delta \phi \simeq H_{\rm inf}/2\pi$ during inflation.
The anisotropic CB can be sizable especially for a smaller decay constant satisfying $f_\phi \sim H_{\rm inf} (\ll 10^{16} \GeV)$.
Note that the monopole abundance required for our scenario can be correspondingly smaller, and it may only be a fraction of DM. Note also that the ALP with $g_{\phi \gamma \gamma} \gtrsim {\cal O}(10^{-12}) \GeV^{-1}$ can be searched for by Fermi-LAT satellite ~\cite{Meyer:2016wrm} and solar axion experiments such as IAXO~\cite{Irastorza:2011gs,Armengaud:2014gea,Armengaud:2019uso}).

In this letter we have pointed out that the recently reported hint for isotropic CB can be naturally explained if the ALP has an effective Hubble-induced mass term through its coupling to the DM energy density. Such an effective mass term can be indeed generated if the ALP is coupled to hidden monopole DM
for the decay constant $f_\phi$ of order the GUT scale. Our scenario applies to a broader range of the ALP mass $m_\phi \lesssim 10^{-28}$\,eV, even smaller than the present Hubble constant.

\vspace{2mm}
%
{\bf Acknowledgments.--}
S.N. acknowledges support from GP-PU at Tohoku University.
The present work is supported by JSPS KAKENHI Grant Numbers
17H02878 (F.T.), 20H01894 (F.T.),
20H05851 (F.T. and M.Y.), JP20K22344 (M.Y.), 
World Premier International Research Center Initiative (WPI Initiative), MEXT, Japan. 
M.Y. was supported by the Leading Initiative for Excellent Young Researchers, MEXT, Japan. 
%

\vspace{1cm}

\bibliography{references}

\end{document}